\documentclass[10pt]{article}
\usepackage{amsfonts}
\usepackage{latexsym}
\usepackage{color, amsmath,amssymb, amsfonts, amstext,amsthm, latexsym, dsfont}
\usepackage{epic,eepic}
\usepackage{graphicx}
\usepackage{longtable}
\usepackage{appendix}
\usepackage{graphicx}
\usepackage{float}
\usepackage[colorlinks=false, linktocpage=true]{hyperref}

\numberwithin{equation}{section}
 \numberwithin{Lem}{section}
 \numberwithin{Defi}{section}
 \numberwithin{Theo}{section}
 \numberwithin{Rem}{section}
  \numberwithin{Coro}{section}
  \numberwithin{Fig}{section}

\title{An Onsager-Machlup approach to the most probable transition pathway for a genetic regulatory network
\footnotetext{\\
Jianyu Hu\\
School of Mathematics and Statistics \& Center for Mathematical Sciences \& Hubei National Center for Applied Mathematics, Huazhong University of Science and Technology, Wuhan
430074, China.\\
E-mail: jianyuhu@hust.edu.cn}
\footnotetext{\\
Xiaoli Chen\\
Department of Mathematics, National University of Singapore, 119076, Singapore.\\
E-mail: xlchen@nus.edu.sg}
\footnotetext{\\Jinqiao Duan\\
Department of Applied Mathematics, Illinois Institute of Technology, Chicago, IL 60616,
USA.\\
E-mail: duan@iit.edu\\
$^*$ Corresponding author}}
\author{Jianyu Hu, Xiaoli Chen$^{*}$ and Jinqiao Duan}

\begin{document}

\maketitle
\begin{abstract}
We investigate a quantitative network of gene expression dynamics describing the competence development in Bacillus subtilis.
First, we introduce an Onsager-Machlup approach to quantify the most probable transition pathway for both excitable and bistable dynamics. Then, we apply a machine learning method to calculate the most probable transition pathway via the Euler-Lagrangian equation. Finally, we analyze how the noise intensity affects the transition phenomena.

\end{abstract}

\noindent\textbf {Lead paragraph:}

\textbf{Gene expression is a typical complex system of high interest. Stochastic effects in gene expression have been shown to generate significant variability. Such effects can result in the transition between metastable states. The key transcription and regulatory factors switch stochastically between the ``on'' (the high concentration) and the ``off'' (the low concentration) states, which correspond to the metastable states.
Thus, quantifying the transition dynamics contributes to the
understanding of the gene expression process. We aim to explore the noise-induced most probable transition pathway of the gene network.
The Onsager-Machlup action function theory provides an approach to quantify the most probable transition pathway, which satisfies the Euler-Lagrange equation. We use the physics informed neural network method to compute the most probable transition pathway. We investigate both excitable and bistable dynamics, and explore how the noise intensity affects the transition phenomena. In excitable dynamics, for small noise intensities, the most probable transition pathways stay around the low concentration state. For larger noise intensities, the transition path can transit to a high concentration and stay in a high concentration for a long time. Therefore, a large noise intensity benefits the expression of comK gene. In bistable dynamics, there is no oscillation for the transition with small noise intensity. The oscillation occurs, when the noise intensity is large. As the noise intensity continuously increases, the number of oscillations increases.
The results may provide a potential guidance for the researchers when they do the experiments. }

\section{Introduction}

Stochastic mechanisms are ubiquitous in biology, physics, geophysics, chemistry, finance, and engineering, and have a profound impact on dynamics.
Random fluctuations result in transitions between metastable states, while deterministic dynamical systems would otherwise be impossible.
Such transitions are responsible for the metastable phenomena observed in systems, such as gene regulatory networks \cite{Acar2008StochasticSA,suel2006excitable,Swain2002IntrinsicAE, Zheng2016TransitionsIA}, regime changes in climate \cite{Rothman2019CharacteristicDO,Wei2021MostPT}, chemical reacting systems \cite{Dykman1994LargeFA,Liu1998StabilityOA}, and fluctuation-activated switching in physics \cite{Bomze2012NoiseinducedCS,Chan2008PathsOF,ma2020precursor}. Therefore, quantifying the transition dynamics contributes to the understanding of such systems.

Gene expression is usually accompanied by transcription of messenger RNA and synthesis of proteins. The key transcription and regulatory factors switch stochastically between the “on” (the high concentration) and the “off”(the low concentration)
state.
In the study of Bacillus subtilis, extensive research developed a MeKS network of molecular interactions that comprise the competence control circuit \cite{grossman1995genetic,hamoen2003controlling}. Stochastic effects in gene expression have been shown to generate significant variability \cite{elowitz2002stochastic,ozbudak2002regulation}. Such effects can result in the transition between metastable states. S\"{u}el et.al.\cite{suel2006excitable}  formulated a mathematical dynamical model to investigate the sample paths of the MeKS gene regulation network. Zhou et.al.\cite{Zhou2021DissectingTC} presented a method based on multiscale reduction technique to identify the underlying stochastic dynamics that prescribes cell-fate transitions. In \cite{chen2019most, wu2018levy}, the authors computed the most likely trajectories using the Fokker–Planck equation. However, the end point of the most likely trajectory depends on the Fokker–Planck equation, which may be not the fixed high concentration state.
It is more meaningful to study the transition between two fixed states.

The Onsager-Machlup action function theory is an effective tool to capture transition behaviors of stochastic dynamical systems, which has been widely investigated in \cite{capitaine1995onsager,fujita1982onsager,ikeda2014stochastic,shepp1992note}.
More precisely, the Onsager-Machlup action functional gives an estimation of the probability of the solution paths of stochastic MeKS gene regulation network on a small tube. The minimum value of the action functional means the largest probability on the tube.
By minimizing the action functional, we can obtain the most probable transition pathway connecting two fixed states, which satisfies the corresponding Euler-Lagrange equation.
We use a neural network method \cite{Hu2021DatadrivenMT,raissi2019physics} to solve the Euler-Lagrange equation and obtain the most probable transition pathway.

In this paper, we investigate the transition of a stochastic MeKS gene regulation network. We compute the most probable transition pathways and tipping times in both excitable and bistable dynamics, using a neural network method to solve the Euler-Lagrange equation through the Onsager-Machlup action function theory.
Moreover, we explore how the noise intensity affects the transition phenomena. We find that noise intensity plays a role in the transition phenomena. In excitable dynamics, we investigate the transition from low concentration to itself as in \cite{suel2006excitable}. We find larger noise intensity benefits the expression of comK gene. In bistable dynamics, we investigate the transition from low concentration to high concentration as in \cite{chen2019most}.
We find a critical noise intensity, corresponding to the occurrence of an oscillation in the most probable transition pathway. As the noise intensity continue increasing, the number of oscillations increases.

This article is arranged as follows. In section 2, we give a brief description of the stochastic MeKS gene regulation network. We introduce an Onsager-Machlup approach to quantify transition paths of this system and use a machine learning method to compute most probable transition pathway. In section 3, we show the most probable transition pathway results for both exitable and bistable dynamics and analyze the effect of the noisy intensity, followed in section 4 by the conclusion.

\section{Model and method}

\subsection{Transition of the stochastic MeKS gene regulation network}
\begin{figure}[htbp]
    \centering
    \includegraphics[scale=0.7]{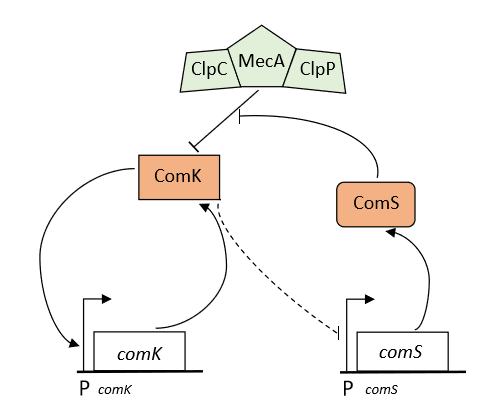}
    \caption{The core competence circuit in B. subtilis (MeKS) \cite{chen2019most}. The $P_{comK}$ ( $P_{comS}$ ) denotes the promoter of comK ( comS ) gene. The ComK protein can activate the expression of comK gene (solid line) and inhibit comS gene to express ComS protein (dotted line). The ComK protein is degraded by the ClpC-MecA-ClpP complex and the ComS protein competitively inhibits ComK protein degradation by the ClpC-MecA-ClpP complex.}
    \label{meksp}
\end{figure}
The core components of the MeKS network are shown in Fig. \ref{meksp} \cite{chen2019most}. The ComK protein can activate the expression of comK gene (positive feedback loop), but is also rapidly degraded by the MecA complex \cite{turgay1998competence}. The ComS protein competitively inhibits ComK protein degradation by the ClpC-MecA-ClpP complex \cite{ogura1999mutational,turgay1998competence}.
However, overexpression of ComK suppresses the expression of comS \cite{hahn1994regulation}.
This implies the existence of an indirect negative feedback loop acting upon ComK, which might affect exit from competence.
Together, MecA complex, comK, comS, the positive and negative feedback loops interacting with each other constitute the MeKS network. To understand how the MeKS network structure determines the dynamics of competence, the enzymatic degradation reactions are assumed to be of the standard Michaelis-Menten form:
\begin{equation}
\begin{split}
&\text { MecA}+\text { ComK } {\stackrel{\gamma_{\pm a}}{\rightleftharpoons}} \text { MecA-ComK } \stackrel{\gamma_{1}}{\rightarrow} \text { MecA }, \nonumber\\
&\text{ MecA }+\text {ComS } \stackrel{\gamma_{\pm b}}{\rightleftharpoons} \text { MecA-ComS } \stackrel{\gamma_{2}}{\rightarrow} \text { MecA }.\nonumber
\end{split}
\end{equation}

By a mathematical analysis, S\"{u}el et.al.\cite{suel2006excitable} formulated a dynamical model to investigate the MeKS gene regulation network
\begin{equation}
\begin{split}
dk&=(a_k+\frac{b_kk^n}{k_0^n+k^n}-\frac{k}{1+k+s})dt,\\
ds&=(\frac{b_s}{1+(k/k_1)^p}-\frac{s}{1+k+s})dt,
\end{split}
\end{equation}
where the symbols $k$ and s represent the concentration of the ComK and ComS proteins, respectively. The parameters $a_k$ and $b_k$ stand for the basal and fully activated rates of ComK protein production, respectively. The parameter $k_0$ denote the concentration of the ComK protein needed for a $50\%$ activation. The Hill coefficients $n$ and $p$ are the cooperativities of the ComK protein auto-activation and ComS protein repression, respectively. ComK inhibits ComS expression by an effective Hill repression function.

Due to random fluctuations, we shall consider a stochastic MeKS gene regulation network
\begin{equation}
\begin{split}\label{meks}
dk&=(a_k+\frac{b_kk^n}{k_0^n+k^n}-\frac{k}{1+k+s})dt+\sigma_1 dW^1(t),\\
ds&=(\frac{b_s}{1+(k/k_1)^p}-\frac{s}{1+k+s})dt+\sigma_2 dW^2(t).
\end{split}
\end{equation}
where $W^1(t)$ and $W^2(t)$ are two independent Brownian motions, $\sigma_1$ and $\sigma_2$ are the noise intensities.

Upon these parameters, the fully activated rates of ComK production $b_k$ and the maximum expression rate of ComS $b_s$ play an important role for the dynamical structures of this model. For different values of $b_k$ and $b_s$, this system has different dynamical structures, which behaviors Hopf bifurcations and saddle-node bifurcations; see \cite{suel2006excitable} for more details. We will consider both excitable and bistable dynamics.

In B. subtilis population, mostly, the ComK protein is expressed by the comK gene at a relative low level, which corresponds to the vegetative state (switch ``off''). Under circumstance of nutrient limitation, a small part of B. subtilis differentiate into the competence state (switch ``on''), in which the ComK protein concentration is high. We mainly consider the transition of the ComK protein concentration.
We want to figure out the most probable transition pathway between the ``off'' state and ``on'' state of stochastic MeKS gene regulation network. The Onsager-Machlup action functional theory provides an approach to quantify the transition paths.

\subsection{An Onsager-Machlup approach}
Now, we briefly introduce the Onsager-Machlup action functional theory and the most probable transition pathway. We shall consider a generalized stochastic differential equation in $\mathbb{R}^d$
\begin{equation}
\begin{split}\label{SDE}
dX(t)=b(X(t))dt+\sigma dW(t), t\in[0,T],
\end{split}
\end{equation}
with initial data $X(0)=x_0 \in \mathbb{R}^d$, where $b:\mathbb{R}^d\rightarrow\mathbb{R}^d$ is a regular function, $\sigma:\mathbb{R}^d\rightarrow\mathbb{R}^{d\times k}$ is a $d\times k$ matrix, and $W$ is a Brownian motion in $\mathbb{R}^k$. Particularly  in the stochastic MeKS gene regulation network \eqref{meks}, $W=(W^1,W^2)^{T}$, the diffusion matrix $\sigma=
\left[
\begin{matrix}
   \sigma_1  &  0\\
   0  & \sigma_2
\end{matrix}
\right]
$, and the vector field $b=(f,g)^T$, where $f(k,s)=a_k+\frac{b_kk^n}{k_0^n+k^n}-\frac{k}{1+k+s}$ and $g(k,s)=\frac{b_s}{1+(k/k_1)^p}-\frac{s}{1+k+s}$.

To investigate transition phenomena, one should estimate the probability of the solution paths on a small tube. The Onsager-Machlup action functional theory of stochastic dynamical system \eqref{SDE} gives an approximation of the probability
\begin{equation}
\begin{split}
\mathbb{P}(\{\|X-z\|_T \leqslant \delta\}) \propto C(\delta,T) \exp \left\{-S(z,\dot{z})\right\},
\end{split}
\end{equation}
where $\delta$ is positive and sufficiently small, $\|\cdot\|_T$ is the uniform norm of the space of all continuous functions in the time interval $[0,T]$, and the Onsager-Machlup action functional is
\begin{equation}
\begin{split}\label{action}
S(z,\dot{z})=\frac{1}{2}\int_0^T\left[\dot{z}-b(z)\right][\sigma\sigma^T]^{-1}[\dot{z}-b(z)]+\nabla \cdot b(z)dt.
\end{split}
\end{equation}
The Onsager-Machlup action functional can be considered as the integral of a Lagrangian
\begin{equation}
\begin{split}
L(z,\dot{z})=\frac{1}{2}\left[\dot{z}-b(z)\right][\sigma\sigma^T]^{-1}[\dot{z}-b(z)]+\frac{1}{2}[\nabla \cdot b(z)].
\end{split}
\end{equation}

The minimum value of the action functional \eqref{action} means the largest probability of the solution paths on a small tube.
We can investigate the most probable transition pathway by minimizing the the Onsager-Machlup action functional.
It captures sample paths of the largest probability around its neighbourhood.
To obtain the most probable transition pathway, we need to find the minimizer of the following optimization problem
\begin{equation}
\begin{split}\label{mini}
\inf_{z \in \bar{C}_{z_1}^{z_2}(0,T)} S(z,\dot{z}),
\end{split}
\end{equation}
where $\bar{C}_{z_1}^{z_2}(0,T)$ is the space of all absolutely continuous functions that start at $z_1$ and end at $z_2$. By the variational principle, the most probable transition pathway from $z_1$ to $z_2$ satisfies the Euler-Lagrange equation
\begin{equation}
\begin{split}\label{EL_eqn}
\frac{d}{dt}\frac{\partial}{\partial \dot{z}}L(z,\dot{z})=\frac{\partial}{\partial z}L(z,\dot{z}),
\end{split}
\end{equation}
with initial state $z(0)=z_1$ and final state $z(T)=z_2$. Particularly in the stochastic MeKS gene regulation network \eqref{meks}, the Euler-Lagrange equation \eqref{EL_eqn} reduces to
\begin{equation}
\begin{split}\label{EL_brf}
\ddot{z}=h(z,\dot{z}),
\end{split}
\end{equation}
where $z=(x,y)^T$ and $h=(h^1,h^2)^T$ such that
\begin{equation}
\begin{split}
h^1(x,y,\dot{x},\dot{y})&=\dot{y}\left(\partial_y f-\frac{\sigma_{1}^{2}}{\sigma_{2}^{2}} \partial_x g\right)+f \partial_{x} f+\frac{\sigma_{1}^{2}}{\sigma_{2}^{2}} g \partial_{x} g+\frac{\sigma_{1}^{2}}{2} \partial_{x} \partial_x f+\frac{\sigma_{1}^{2}}{2} \partial_x \partial_y g, \\
h^2(x,y,\dot{x},\dot{y})&=\dot{x}\left(\partial_{x} g-\frac{\sigma_{2}^{2}}{\sigma_{1}^{2}} \partial y f\right)+\frac{\sigma_{2}^{2}}{\sigma_{1}^{2}} f \partial_y f+g \partial_{y} g+\frac{\sigma_{2}^{2}}{2} \partial_y \partial_x f+\frac{\sigma_{2}^{2}}{2} \partial_y \partial_y g.
\end{split}
\end{equation}
Here, the components of the vector field are $f(k,s)=a_k+\frac{b_kk^n}{k_0^n+k^n}-\frac{k}{1+k+s}$ and $g(k,s)=\frac{b_s}{1+(k/k_1)^p}-\frac{s}{1+k+s}$. In next subsection, we will introduce the physics informed neural networks method to numerically calculate the Euler-Lagrange equation \eqref{EL_brf} to explore the most probable dynamics of the stochastic MeKS gene regulation network.

\subsection{A machine learning approach to compute most probable transition pathway}
The physics informed neural networks method is successfully applied to compute the most transition pathway, which is an effective method, even for high dimensional systems. The idea is to construct a neural network for the most probable transition pathway and consider the governing equation as a part of the loss function.

We approximate the solution to the Euler-Lagrange equation \eqref{EL_brf} from a set of training data, which consist of residual data, initial and final data.
Denote the residual data as $\{(t_j,\ddot{z}-h(z,\dot{z}))\}_{j=1}^m$, where $t_j\in(0,T)$, and the initial and final data $\{(0,z_1),(T,z_2)\}$, where $m$ is the number of the residual data. Therefore the the empirical loss function is
\begin{equation}
\begin{split}\label{loss}
\operatorname{Loss}_{m}\left(z,\lambda\right)=\frac{1}{m}\sum\limits_{j=1}^m\|\ddot{z}(t_j)-h(z(t_j),\dot{z}(t_j))\|^2+ \frac{\lambda}{2}(\|z(0)-z_1\|^2+\|z(T)-z_2\|^2),
\end{split}
\end{equation}
where $\lambda$ is a positive constant, which is used to balance the residual loss and boundary loss.


\section{Analysis of the most probable transition pathway}
In this section, we analyze the most probable transition pathway of the stochastic MeKS gene regulation network  \eqref{meks}. We consider both
excitable and bistable dynamics, which corresponds to different parameters choice. We study the effect of noise intensity to the most probable transition pathway. For convenience, we take $\sigma_1=\sigma_2=\sigma$.

In the following computing, we construct full connected neural networks to compute the most transition pathway by optimizing the empirical loss function  \eqref{loss}. All the neural networks have 4 hidden layers and 20 neurons per layer, with $\operatorname{tanh}$ activation function. The weights are initialized with truncated normal distributions and the biases are initialized as zero.
The Adam optimizer with a learning rate of $10^{-4}$ is used to train the loss function. The number of residual points for evaluating the Euler-Lagrange equation \eqref{EL_brf} is $m=501$.

\subsection{Transition in  excitable dynamics}
\begin{figure}
\begin{minipage}[]{0.5 \textwidth}
 \leftline{~~~~~~~\tiny\textbf{(a)}}
\centerline{\includegraphics[width=6cm,height=5cm]{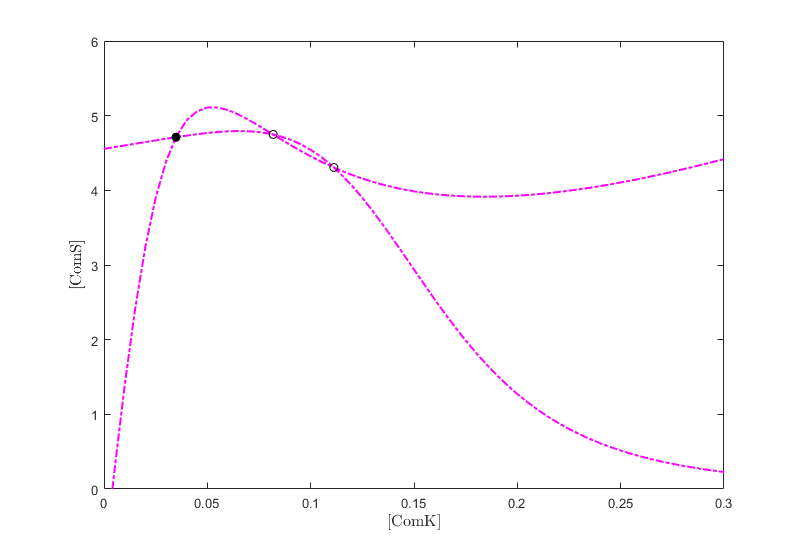}}
\end{minipage}
\hfill
\begin{minipage}[]{0.5 \textwidth}
 \leftline{~~~~~~~\tiny\textbf{(b)}}
\centerline{\includegraphics[width=6cm,height=5cm]{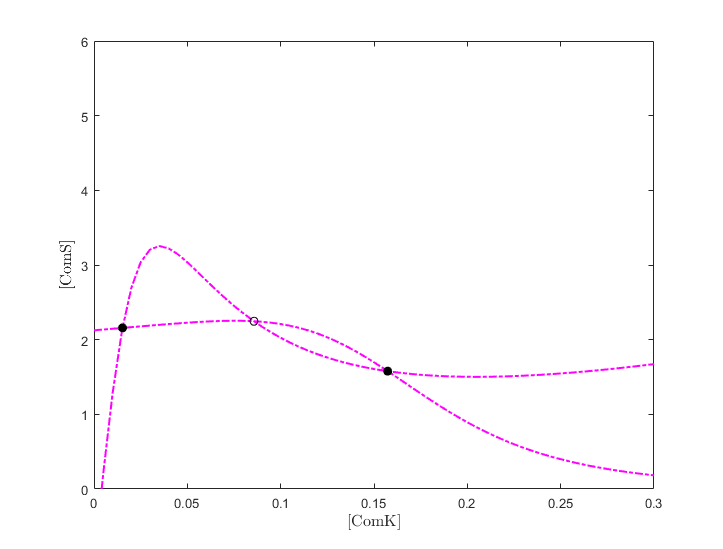}}
\end{minipage}
\caption{\textbf{Nulllines for the MeKS model}. The pink lines are the nulllines of the corresponding deterministic system. (a) Excitability: $b_k=0.07$ and $b_s=0.82$. The intersection points are equilibria corresponding to a stable black solid point and two unstable hollow points. (b) Bistability: $b_k=0.14$ and $b_s=0.68$. The intersection points are equilibria corresponding to two stable black solid points and an unstable hollow point.}
\label{nullline}
\end{figure}

In this subsection, we consider the transition of the excitable dynamics. We chose $a_k=0.004$, $k_0=0.2$, $k_1=0.222$, $n=2$ and $p=5$ in equation \eqref{meks}. The bifurcation parameters are chosen as $b_k=0.07$ and $b_s=0.82$. In this situation, the system behaves excitable dynamics.
We set the drift vector field equal to zero, and obtain the dynamical nulllines of the gene expression system; see Fig.\ref{nullline}(a). There are three equilibria corresponding to the deterministic system: (i) the stable state $(k_1,s_1)\approx (0.03485,4.7117)$; (ii) the unstable states $(k_2,s_2)\approx (0.08184,4.7484)$ and $(k_3,s_3)\approx (0.11121,4.3058)$.

We compute the most probable transition pathway through minimizing the loss function \eqref{loss} with $\lambda=100$. The transition time $T = 10$. Inspired by the paper \cite{suel2006excitable}, we consider the transition from low  concentration $(k_1,s_1)=(0.03485,4.7117)$ to itself. Moreover, we study the effect of the noise intensity to the most probable dynamics.

To show the speed of transcription, we define a tipping time to characterise this dynamics. The tipping time is time firstly passing the saddle point. Mathematically, we define a general tipping time $T^{k}(x)$ associated to the $k$th-component of a point $p=(p_1,...,p_d)$ for the transition path $x$ with $x_k(0)<p_k$ as follows
$$
T^{k} = \inf\limits_{t}\{t:x_k(t)-p_k>0\}.
$$

\noindent\textbf{The effect of the noise intensity}
\begin{figure}
\begin{minipage}[]{0.3 \textwidth}
 \leftline{~~~~~~~\tiny\textbf{(a)}}
\centerline{\includegraphics[width=4.8cm,height=4.4cm]{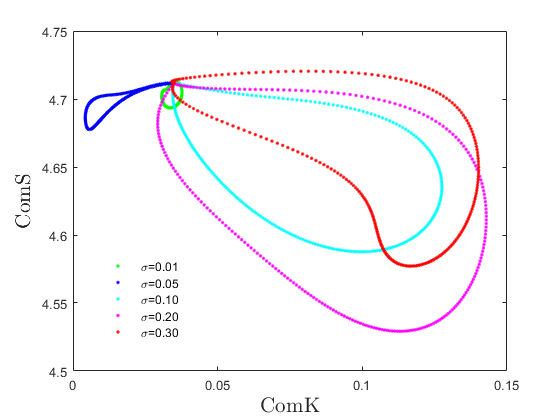}}
\end{minipage}
\hfill
\begin{minipage}[]{0.3 \textwidth}
 \leftline{~~~~~~~\tiny\textbf{(b)}}
\centerline{\includegraphics[width=4.8cm,height=4.4cm]{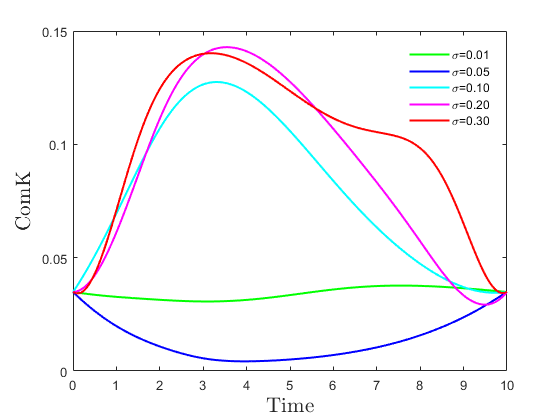}}
\end{minipage}
\hfill
\begin{minipage}[]{0.3 \textwidth}
 \leftline{~~~~~~~\tiny\textbf{(c)}}
\centerline{\includegraphics[width=4.8cm,height=4.4cm]{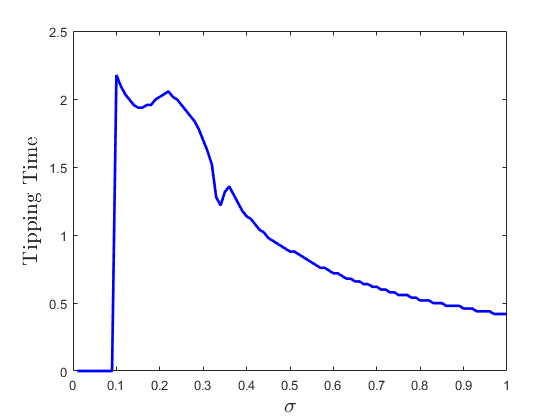}}
\end{minipage}
\caption{\textbf{The most probable transition pathways for excitable dynamics with different noise intensities}. The noise intensity $\sigma \in (0,1]$ and transition time $T=10$. (a) The most probable transition pathways in the ComK-ComS plane. (b) The most probable transition pathways in the ComK space with respect to time. (c) Tipping time with respect to the noisy intensity. }
\label{meks_excitable}
\end{figure}

In the stochastic MeKS gene regulation network, the noise intensity plays an important role in the system. It represents the effect of the cell environment to the gene expression process. Thus, we try to investigate the effects of the noise intensity on transition phenomena.

We take different noise intensities $\sigma=0.01* i$ for $i=1,2,...,100$ and compute the corresponding most probable transition pathways.
The results are shown in Fig.\ref{meks_excitable}. The tipping times with respect to the noisy intensity are presented in Fig.\ref{meks_excitable}(c). There is critical noise intensity $\sigma^{*}=0.1$, which corresponds to the possibility of the transition to high concentration. When $\sigma\geq\sigma^{*}$, the larger noise intensity roughly results in smaller tipping time.
The most probable transition pathways for different noise intensities in the ComK-ComS plane are presented in Fig.\ref{meks_excitable}(a) and the ComK component with respect to time is shown in Fig.\ref{meks_excitable}(b).
For small noise intensities $(\sigma<\sigma^{*})$, the most probable transition pathways stay around the low concentration state. Hence, a small noise intensity is not in favor of the expression of comK gene. For lager noise intensities $(\sigma\geq\sigma^{*})$, the transition path can transit to high concentration and stay in high concentration for a long time. Therefore, a large noise intensity benefits the expression of comK gene.

\noindent\textbf{Perturbation effects for the initial state}

\begin{figure}
\begin{minipage}[]{0.5 \textwidth}
 \leftline{~~~~~~~\tiny\textbf{(a)}}
\centerline{\includegraphics[width=6.4cm,height=5.4cm]{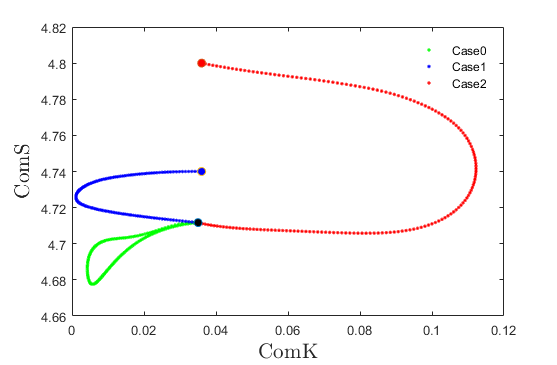}}
\end{minipage}
\hfill
\begin{minipage}[]{0.5 \textwidth}
 \leftline{~~~~~~~\tiny\textbf{(b)}}
\centerline{\includegraphics[width=6.4cm,height=5.4cm]{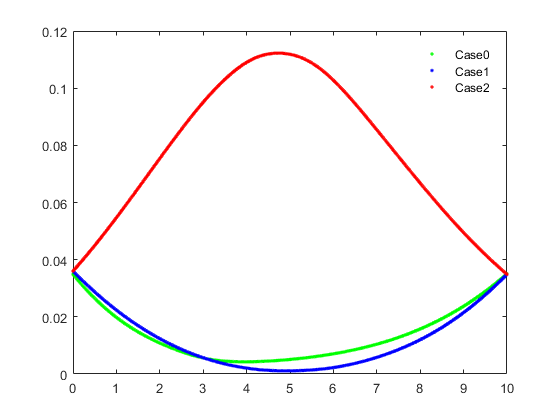}}
\end{minipage}
\hfill
\begin{minipage}[]{0.5 \textwidth}
 \leftline{~~~~~~~\tiny\textbf{(c)}}
\centerline{\includegraphics[width=6.4cm,height=5.4cm]{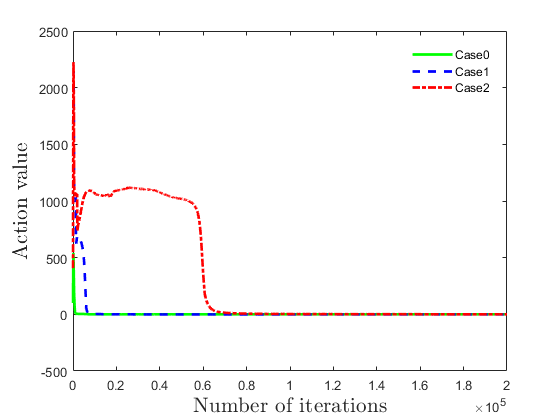}}
\end{minipage}
\hfill
\begin{minipage}[]{0.5 \textwidth}
 \leftline{~~~~~~~\tiny\textbf{(d)}}
\centerline{\includegraphics[width=6.4cm,height=5.4cm]{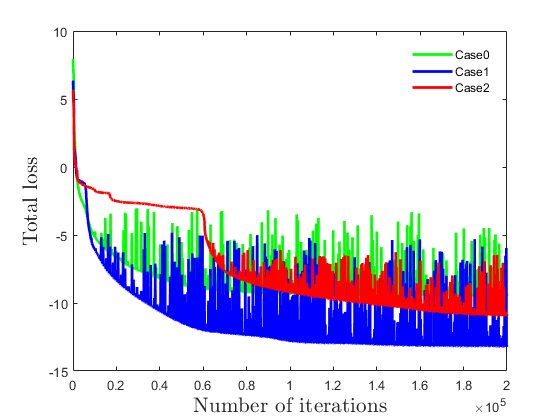}}
\end{minipage}
\caption{\textbf{Perturbation effects to the initial state for excitable dynamics}. The initial state $(k,s)$ are chosen for three cases. Case0 (no perturbation): $(k,s)=(0.03485,4.7117)$; Case1 (small perturbation):  $(k,s)=(0.036,4.74)$; Case2 (large perturbation): $(k,s)=(0.036,4.8)$. (a) The most probable transition pathwayS in the ComK-ComS plane. (b) The most probable transition pathways of ComK with respect to time. (c) Thhe value of the Onsager-Machlup action functional as the iteration of the optimizer progresses. (d) The logarithm of total loss as the iteration of the optimizer progresses.}
\label{meks_excitable_perturbation}
\end{figure}

For $\sigma=0.05$, the most probable transition pathways stay around the low concentration state. We consider perturbations for the initial states and figure out whether it can transit to high concentration state. We study three cases for the initial state. Case0 (no perturbation): $(k,s)=(0.03485,4.7117)$; Case1 (small perturbation):  $(k,s)=(0.036,4.74)$; Case2 (big perturbation): $(k,s)=(0.036,4.8)$. The results are shown in Fig.\ref{meks_excitable_perturbation}.

To show the efficiency of the neural network, we present the value of the Onsager-Machlup action functional and the logarithm of total loss with respect to the the iteration of the optimizer progresses, which are shown in Fig.\ref{meks_excitable_perturbation}(c) and Fig.\ref{meks_excitable_perturbation}(d). The values of Onsager-Machlup action functional for all three cases decrease rapidly and then tend to be stable. The total loss for all three cases reaches about $10^{-9}$, which shows the efficiency of our computing. The most probable transition pathway the stochastic MeKS model in the ComK-ComS plane is presented in Fig.\ref{meks_excitable_perturbation}(a) and the ComK component with respect to the time is shown in Fig.\ref{meks_excitable_perturbation}(b). For small perturbation (Case1), the most probable transition pathway also stays around the low concentration state. But for large perturbation (Case2), the path can transit high concentration state.

\subsection{Transition in bistable dynamics}
In this subsection, we consider the transition in bistable dynamics. The parameters of the excitable region for our numerical experiments, namely $a_k=0.004$, $k_0=0.2$, $n=2$, $p=5$, $b_k=0.14$ and $b_s=0.68$. In this situation, the system behaviors bistable dynamics. From the nullclines in Fig.\ref{nullline}(b), there are three equilibria corresponding to the deterministic system:  (i) the stable states $(k_1,s_1)\approx (0.015262, 2.1574)$ and $(k_3,s_3)\approx (0.15732, 1.5781)$; (ii) the unstable state $(k_2,s_2)\approx (0.08568, 2.2469)$.

We compute the most probable transition pathway through minimizing the loss function \eqref{loss} with $\lambda=100$. The transition time $T = 5$. We consider the transition from low  concentration $(k_1,s_1)=(0.015262, 2.1574)$ to high concentration $(k_3,s_3)=(0.15732, 1.5781)$. Moreover, we study the effect of the noise intensity to the most probable dynamics.

\noindent \textbf{Effects of the noise intensity}

\begin{figure}
\begin{minipage}[]{0.3 \textwidth}
 \leftline{~~~~~~~\tiny\textbf{(a)}}
\centerline{\includegraphics[width=4.8cm,height=4.4cm]{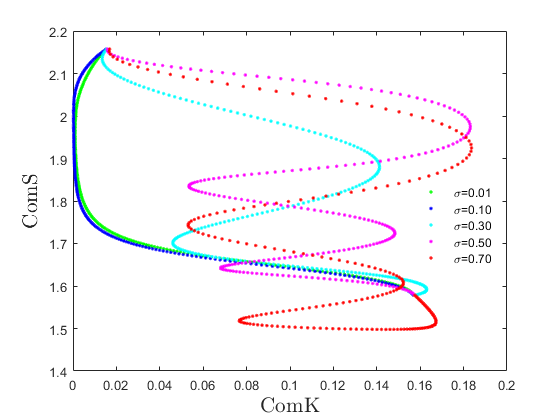}}
\end{minipage}
\hfill
\begin{minipage}[]{0.3 \textwidth}
 \leftline{~~~~~~~\tiny\textbf{(b)}}
\centerline{\includegraphics[width=4.8cm,height=4.4cm]{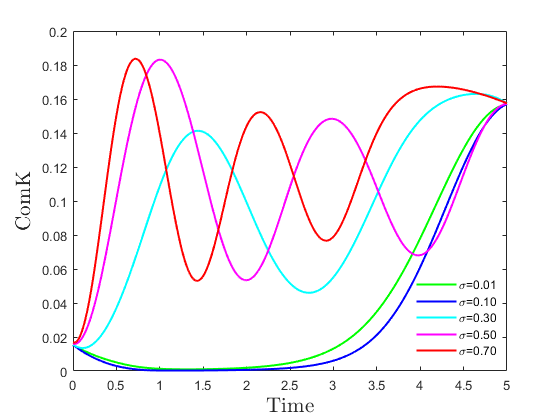}}
\end{minipage}
\hfill
\begin{minipage}[]{0.3 \textwidth}
 \leftline{~~~~~~~\tiny\textbf{(c)}}
\centerline{\includegraphics[width=4.8cm,height=4.4cm]{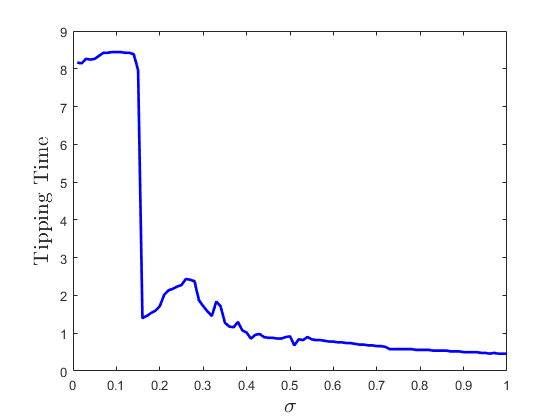}}
\end{minipage}
\caption{\textbf{The most probable transition pathways for bistable dynamics with different noise intensities}. The noise intensity $\sigma\in (0,1]$ and transition time $T=5$. (a) The most probable transition pathways in the ComK-ComS plane. (b) The most probable transition pathways in the ComK space with respect to time. (c) Tipping time with respect to the noisy intensity. }
\label{meks_bistable}
\end{figure}

Similar to excitable case, we take different noise intensities $\sigma=0.01 * i$ for $i=1,2,...,100$. We compute the most probable transition pathways and tipping times for different noise intensities. The results are shown in Fig.\ref{meks_bistable}. The tipping times with respect to the noisy intensity are presented in Fig.\ref{meks_bistable}(c). There is critical noise intensity $\sigma^{+}=0.16$, which corresponds to the occurrence of oscillation. The most probable transition pathways for different noise intensities in the ComK-ComS plane are presented in Fig.\ref{meks_bistable}(a) and the ComK component with respect to time is shown in Fig.\ref{meks_bistable}(b).
When the noise intensity is relatively small $(\sigma<\sigma^{+})$, there is no oscillation for the transition; see the  green and blue curves. In this situation, the trajectory stays in low concentration for a long time and then transits to high concentration, which is similar to the result in \cite{chen2019most}.
When the noise intensity is large $(\sigma\geq\sigma^{+})$, the oscillation occurs. As the noise intensity continuously increases, the number of oscillation increases.
The results may give a guidance for the researchers to choose a lager noise intensity when they do the experiment.

\section{Conclusion}
In this paper, we investigate the transition in the stochastic gene regulation network for both excitable and bistable dynamics.
We introduce the Onsager-Maclup action functional theory to quantify the transition and obtain the most probable transition pathway. We construct a neural network to compute the most probable transition pathway. Moreover, we compute the tipping time, and study the effects of the noise intensity which plays a role in the transition phenomena.

In excitable dynamics, for small noise intensities, the most probable transition pathways stay around the low concentration state. In this situation, we further investigate the perturbation effects for the initial state. For small perturbation, the most probable transition pathway also stays around the low concentration state. But for large perturbations, the path can transit the high concentration state. For larger noise intensities, the transition path can transit to the high concentration state. A large noise intensity is in favor of the expression of comK gene.

In bistable dynamics, there is no oscillation for the transition with small noise intensity. The oscillation occurs, when the noise intensity is large. As the noise intensity continuously increases, the number of oscillation increases.

\section*{Acknowledgements}
We would like to thank Prof. Renming Song for helpful talks and thank Ying Chao, Yanjie Zhang, Qi Zhang, Jinayu Chen, Pingyuan Wei, Yuanfei Huang, Qiao Huang, Wei Wei and Ao Zhang for helpful discussions. This work was partly supported by NSFC grants 11771449 and 11531006.

\bibliographystyle{abbrv}
\bibliography{Ref}

\end{document}